\documentclass[sigconf]{acmart}
\usepackage{enumitem}
\usepackage{lipsum}  
\usepackage{subcaption}
\usepackage{graphicx}
\usepackage{tabularx}
\usepackage{booktabs}
\usepackage{soul}
\usepackage{algorithm}
\usepackage{algpseudocode}
\usepackage{multirow}
\usepackage{diagbox}

\usepackage{amsthm}  % For theorem-like environments
\usepackage{amsmath}
\usepackage{cleveref}
\newtheorem{assumption}{Assumption}

\newcommand\dhist{\textit{TKHist}}
\newcommand\card{CE}
\newcommand\topk{top-k}

\algnewcommand{\LeftCommentNew}[1]{\Statex \hspace{-0.2cm} \(\triangleright\) #1}
\algnewcommand{\LeftComment}[1]{\Statex \hspace{0.4cm} \(\triangleright\) #1}

\AtBeginDocument{%
  }

\copyrightyear{2025} 
\acmYear{2025} 
\setcopyright{acmlicensed}
\acmConference[CIKM '25]{Proceedings of the 34th
 ACM International Conference on Information and Knowledge
 Management}{November 10--14, 2025}{Seoul, Republic of Korea}
 \acmBooktitle{Proceedings of the 34th ACM International Conference on
 Information and Knowledge Management (CIKM '25), November 10--14, 2025,
 Seoul, Republic of Korea}
 \acmDOI{10.1145/3746252.3761219}
 \acmISBN{979-8-4007-2040-6/2025/11}
\begin{document}

\title{\dhist: Cardinality Estimation for Join Queries via Histograms with Dominant Attribute Correlation Finding }

\author{Renrui Li}
% \authornote{Both authors contributed equally to this research.}
\affiliation{%
  \institution{Soochow University}
  \department{School of Computer Science and Technology}
  \city{Suzhou}
  \state{Jiangsu}
  \country{China}
}
\email{rrli@stu.suda.edu.cn}
\orcid{0009-0000-6572-5924}

\author{Qingzhi Ma}
\authornote{Corresponding author.}
% \authornote{Both authors contributed equally to this research.}
\affiliation{%
  \institution{Soochow University}
  \department{School of Computer Science and Technology}
  \city{Suzhou}
  \state{Jiangsu}
  \country{China}
}
\email{qzma@suda.edu.cn}
\orcid{0000-0003-2418-090X}

\author{Jiajie Xu}
\affiliation{%
  \institution{Soochow University}
  \department{School of Computer Science and Technology}
  \city{Suzhou}
  \state{Jiangsu}
  \country{China}
}
\email{xujj@suda.edu.cn}
\orcid{0000-0001-8227-8636}

\author{Lei Zhao}
\affiliation{%
  \institution{Soochow University}
  \department{School of Computer Science and Technology}
  \city{Suzhou}
  \state{Jiangsu}
  \country{China}
}
\email{zhaol@suda.edu.cn}
\orcid{0000-0002-5123-9279}

\author{An Liu}
\affiliation{%
  \institution{Soochow University}
  \department{School of Computer Science and Technology}
  \city{Suzhou}
  \state{Jiangsu}
  \country{China}
}
\email{anliu@suda.edu.cn}
\orcid{0000-0002-6368-576X}

\begin{abstract}
Cardinality estimation has long been crucial for cost-based database optimizers in identifying optimal query execution plans, attracting significant attention over the past decades.
While recent advancements have significantly improved the accuracy of multi-table join query estimations, these methods introduce challenges such as higher space overhead, increased latency, and greater complexity, especially when integrated with the binary join framework.
In this paper, we introduce a novel cardinality estimation method named \dhist{}, which addresses these challenges by relaxing the uniformity assumption in histograms.
\dhist{} captures bin-wise non-uniformity information, enabling accurate cardinality estimation for join queries without filter predicates.
Furthermore, we explore the attribute independent assumption, which can lead to significant over-estimation rather than under-estimation in multi-table join queries.
To address this issue, we propose the dominating join path correlation discovery algorithm to highlight and manage correlations between join keys and filter predicates.
Our extensive experiments on popular benchmarks demonstrate that \dhist{} reduces error variance by 2-3 orders of magnitude compared to SOTA methods, while maintaining comparable or lower memory usage.
\end{abstract}




\maketitle
\vspace{-0.4cm}
\section{Introduction}
Cardinality estimation (\card{}) is a crucial component of cost-based query optimizers in modern database systems. 
Its goal is to accurately and efficiently estimate the size of relevant tuples or the selectivity of a query, without requiring actual database execution, thereby improving the database performance.

The well-known Selinger model \cite{selinger1979access}, which relies on the assumptions of attribute independence and join-key uniformity, has been widely adopted in various database systems due to its low latency and efficient query plan search capabilities.
Join Histogram \cite{adellera} mitigates the join-key uniformity assumption, providing improved accuracy for multi-table join queries. 
However, real-world datasets frequently deviate from these strong assumptions, leading to significant errors in join cardinality estimation. 
For complex queries, estimated join size errors can reach up to 4 orders of magnitude and grow exponentially with the number of tables in join queries, severely degrading the quality of execution plans.

With the prosperity of artificial intelligence, query-driven and data-driven methods have emerged.
Query-driven methods, such as those proposed in \cite{kipf2018learned,liu2021fauce, sun2019end}, employ supervised learning to model the mapping between query statements and their corresponding cardinality, achieving high accuracy. 
However, these methods rely heavily on predefined query templates, which limits their applicability to ad hoc or previously unseen queries.
Data-driven methods \cite{getoor2001selectivity, hilprecht2020deepdb, tzoumas2011lightweight, wu2020bayescard, yang2020neurocard, zhu2021flat} typically construct a denormalized representation of join tables to model the underlying high-dimensional data distribution.
Although these methods are capable of producing near-optimal execution plans, they often incur large model sizes and pose significant integration challenges for database systems that adopt a binary join framework.
Moreover, some studies indicate that cardinality over-estimation tends to provide more robust execution plans than cardinality under-estimation \cite{abo2017shannon, atserias2013size, cai2019pessimistic, hertzschuch2021simplicity,leis2015good}.
% Consequently, upper-bound based methods, such as PessEst \cite{cai2019pessimistic} and FactorJoin \cite{wu2023factorjoin}, have emerged. 
PessEst\cite{cai2019pessimistic} utilizes randomized hashing and data sketching to produce a tightened upper bound for join queries. 
Building on this approach, FactorJoin\cite{wu2023factorjoin} overestimates cardinality using single-table statistics and the PGM inference algorithm, providing state-of-the-art performance.
These efforts aim to design effective \card{} methods by addressing or mitigating the aforementioned strong assumptions.

In general, an effective \card{} method should satisfy several key criteria \cite{wu2023factorjoin, kim2024asm}: It should provide sufficient accuracy to identify high-quality query plans while ensuring efficiency across critical metrics, including low latency, minimal training time, low space overhead, scalability with the number of join tables, and update support.
Despite the variety of existing \card{} solutions, few approaches fully satisfy all the aforementioned performance criteria.

\vspace{-0.1cm}
\subsection*{Motivations}
Here, we investigate two interesting phenomena for join queries only, which inspire our study.
\subsubsection*{\textbf{[1]Error Amplification in Binary Join Framework}}
Most database systems rely on the binary join framework, which leads to query errors accumulating and increasing exponentially with the number of tables in a join query. 
% (In \Cref{sec:pure:cardinality}, we will compare the performance of Join Histogram, FactorJoin and DeepDB\cite{hilprecht2020deepdb} for queries without any filter predicates based on the STATS dataset.)
As a learned method, DeepDB delivers relatively accurate estimates, but often underestimates.
While Join Histogram underestimates by up to 4 orders of magnitude.
FactorJoin method overestimates the join results more tightly. 
As the number of tables in join queries increases, errors escalate exponentially.
The trend of increasing errors appears inevitable given the current research within the binary join framework.
% Therefore, a central question in our study is: \textbf{Can we achieve unbiased pure cardinality estimation for join queries within the binary join framework?}
% \textcolor{red}{
% Therefore, a central question in our study is: \textbf{Can we achieve accurate cardinality estimation for pure join queries within the binary join framework?}
% }
Therefore, a central question in our study is: \textbf{Can we achieve accurate cardinality estimation for pure join queries within the binary join framework?}
% Here, pure join queries mean queries without filter predicates.
Here, pure join queries mean queries without filter predicates, which allow a clearer evaluation of a \card{} method’s ability to estimate the cardinality of join operations.

% \begin{figure}[htb!]
%     \centering
%     \includegraphics[width=0.9\linewidth]{fig/dominating_proportion_1.pdf}
%     % \Description{top k }
%     \caption{Significance of top $k$ join paths. The dataset is STATS, and the longest query is 
%     $user\bowtie badges\bowtie comments \bowtie votes \bowtie posts$
%     }
%     \label{fig:top:k}
% \end{figure}

% Additionally, we observe that the attribute independency assumption can cause both under-estimation and over-estimation of join cardinality, a phenomenon not previously discovered(refer to \Cref{sec:multi:table:stats} for more details). 
% This is largely due to the correlation between join keys and filter predicates.
% The uniformity assumption on join keys under the binary framework usually neglects the significance of dominating (non-uniform) join keys, which becomes critical with an increasing number of join tables.

% \subsubsection*{\textbf{[2]Significance of Attribute Correlation in \topk{} Join Paths}}
\subsubsection*{\textbf{[2]Significance of Attribute Correlation in Dominant Join Paths}}
Additionally, the uniformity assumption in histograms can lead to significant errors as the number of joined tables increases. 
This phenomenon primarily stems from the generation of join paths. 
% \textcolor{red}{In this study, a join path means a special equi-join template.} 
In this study, a join path means a set of primary/foreign keys that satisfy the equi-join conditions.
For example, $T1.id=T2.id=1$  represents a join path in the join of two tables.
A dominant join path refers to a join path that contributes a substantial portion of the final join results.
% \textcolor{red}{A dominant join path refers to a join path that contributes a substantial portion of the final join results.}
% \textcolor{red}{join key or join path}
% The uniformity assumption on join keys typically neglects the importance of dominant non-uniform join keys (we use \topk{} join keys in this study) – a factor that becomes increasingly critical with more joined tables. 
% We demonstrate this behavior through the following analysis.
The uniformity assumption on join paths typically neglects the importance of dominant join paths, a factor that becomes increasingly critical with more joined tables. 
We demonstrate this behavior through the following analysis.

\begin{figure}[htb!]
    \centering
    \includegraphics[width=0.8\linewidth]{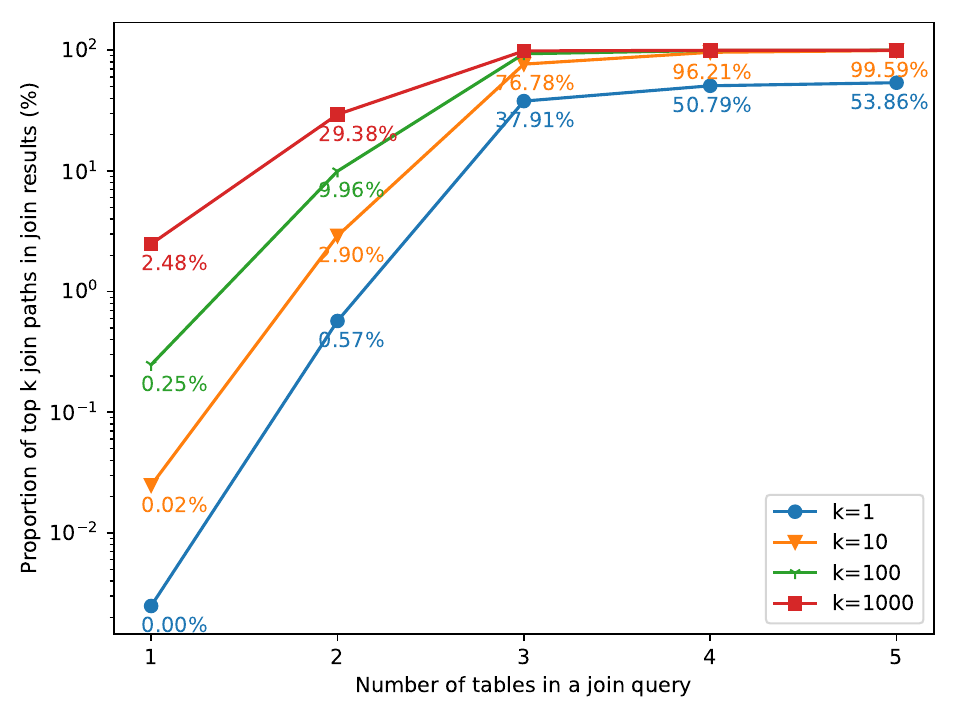}
    \caption{Proportion of top $k$ join paths in join results. The dataset is STATS, and the join query template is 
    $user\bowtie badges\bowtie comments \bowtie votes \bowtie posts$.
    }
    \label{fig:top:k}
    % \vspace{-0.3cm}
\end{figure}

% \Cref{fig:top:k} illustrates the proportion of final join results contributed by \topk{} join paths for a representative STATS-CEB dataset query, where $k$ ranges from 1 to 1000. 
\Cref{fig:top:k} illustrates the proportion of final join results contributed by \topk{} join paths for a representative query in STATS dataset, where $k$ ranges from 1 to 1000. 
As shown, the contribution of \topk{} join paths grows substantially with additional tables. 
In the 5-table join query, the top 10 join paths account for 99.59\% of all results, making them dominant in the final join result.
% When filter predicates fail to effectively exclude elements within these \topk{} join paths, estimates become prone to significant overestimation. 
When filter predicates fail to effectively exclude elements within these dominant join paths, estimates become prone to significant overestimation. 
This observation leads to our second key research question: \textbf{Can we improve query estimation accuracy by explicitly modeling the relationship between join paths and filter predicates?}

\subsection*{Our Approach}
This study aims to establish a new framework for accurate cardinality estimation of join queries using single-table statistics.
Our approach leverages 1D and 2D histograms–widely adopted in database systems for their minimal maintenance requirements and update efficiency.
The framework has two key components:
First, to relax the uniformity assumption, we introduce a novel histogram structure called 
{\dhist{} (\textbf{Hist}ograms with \textbf{T}op-\textbf{K} items)}, 
which accurately captures the non-uniformity information within the histograms. 
Second, we highlight that the correlation between join keys and filter predicates is more significant than correlations between filter predicates.
Building on this key finding, we design a framework to accurately maintain the correlation between the join keys in \topk{} join paths and the filter predicates.

\subsection*{Contributions}
This paper represents the design and implementation of \dhist, a \card{} framework for join queries supporting aforementioned performance criteria by utilizing a pre-built prior state, while maintaining the practical benefits of histogram-based approaches.
It offers:
\begin{itemize}[noitemsep,wide=0pt, leftmargin=\dimexpr\labelwidth + 2\labelsep\relax]
\item  \textbf{Superior Accuracy:}  Reduces error variance by 2-3 orders of magnitude for join queries without predicates.
\item \textbf{Correlation-Aware Estimation:} Dominant join path discovery algorithm that captures predicate-path relationships
% , enabling additional error reduction.
\item  \textbf{Efficiency:} Comparable or shorter query response times with significantly smaller memory footprints, short state-building times, low maintenance, and low update overheads.
% \item High accuracy for join queries, reducing error variance by 2-3 orders of magnitude.
% % \item Comparable or shorter query response times.
% % \item \textcolor{red}{Comparable or shorter query response times.}
% \item {Comparable or shorter query response times.}
% \item Significantly smaller memory footprints.
% \item Short state-building times.
% % \item \textcolor{red}{Short state-building times.}
% \item Low maintenance and update overheads.
\end{itemize}
% \vspace{0.4cm}
% \vspace{0.2cm}
% The paper will:
% \begin{itemize}[noitemsep,wide=0pt, leftmargin=\dimexpr\labelwidth + 2\labelsep\relax]
% \item show how to design and build \dhist{} based on traditional histograms and how to use them for join queries,
% \item analyze the sensitivity and stress \dhist's performance on key parameters, such as top $k$ dominating items to capture non-uniformity within bins, bin size selection to trade-off accuracy and space overheads, etc.
% \item perform a comprehensive performance evaluation of \dhist{}, comparing it against state-of-the-art \card{} methods, including DeepDB, FACE, FactorJoin, etc, using queries based on the STATS-CEB, TPC-DS benchmarks.
The paper will:
\begin{itemize}[noitemsep,wide=0pt, leftmargin=\dimexpr\labelwidth + 2\labelsep\relax]
\item Demonstrate the design and construction of \dhist{} based on traditional histograms
% , and its application to join queries 
(see \Cref{sec:frame}).
\item Analyze \dhist's sensitivity to key parameters, including top $k$ dominating items to capture bin-wise non-uniformity, and bin size to balance accuracy and space overheads (see \Cref{sec:impl}).
\item Conduct a comprehensive performance evaluation of \dhist{}, comparing it with state-of-the-art \card{} methods such as DeepDB, BayesCard, and FactorJoin, using queries from the STATS-CEB and IMDB/JOB-light benchmarks (see \Cref{sec:experiment}).
\end{itemize}

% This study proposes a new \card{} framework to produce unbiased cardinality estimation for join queries based on single-table statistics. 
% The structure of this paper is  as follows:
% Section 2 provides background information. Section 3-4 introduces \dhist{} and its design and implementation details.
% Experimental results are summarized in Section 5. We briefly review related literature in Section 6 and Section 7 concludes the paper.

\section{TKHist Framework for Join Queries}
\label{sec:frame}
This section introduces the high-level design and implementation of \dhist.
% First, we provide an overview of the workflow, including state building and online query answering. 
First, \Cref{sec:workflow} provides an overview of the overall workflow. 
Next, \Cref{sec:dim:reduction} presents the dimension reduction strategy for single tables, which helps reduce the complexity of the data structure.
Subsequently, \Cref{sec:data:structure} introduces the core histogram data structure of \dhist, designed to accurately capture non-uniformity information.
Finally, \Cref{sec:join:path:processing} applies \dhist{} to join query processing under the binary join framework, decomposing joins into paths for efficient cardinality estimation.
% Finally, \Cref{sec:join:path:processing} applies \dhist{} to join query processing under the binary join framework, decomposing joins into paths for efficient cardinality estimation.
% Finally, \Cref{sec:join:path:processing} applies \dhist{} to join query processing under the binary-join framework, decomposing joins into paths for efficient estimation.
The notation for parameters in this section is summarized in \Cref{tab:notation}.

% \subsection{Workflow Overview}
% \label{sec:workflow}
% \begin{figure}[ht!]
%     \centering
%     \includegraphics[width=\linewidth]{fig/workflow1.png}
%     \caption{\dhist{} workflow}
%     \label{fig:workflow}
% \end{figure}
\subsection{Workflow Overview}
\label{sec:workflow}
\begin{figure}[ht!]
    \centering
    \includegraphics[width=0.8\linewidth]
    {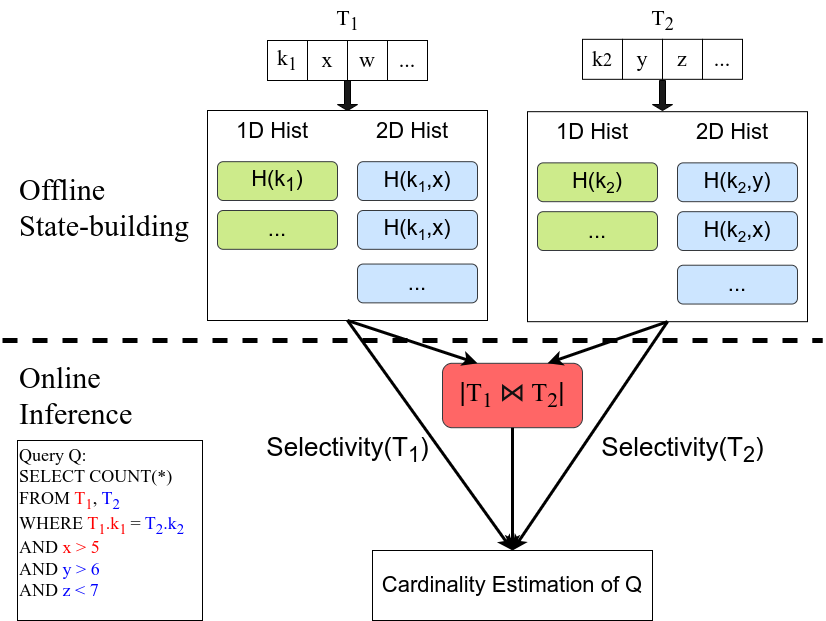}
    \caption{\dhist{} workflow.}
    \label{fig:workflow}
    \vspace{-0.2cm}
\end{figure}

The \dhist{} workflow consists of an offline state-building phase and an online inference phase. Consider a two-table join query example as shown in \Cref{fig:workflow}, where table $T_1$ joins $T_2$ on $T_1.k_1=T_2.k_2$, with three range predicates on non-key attributes $x, y$, and $z$.
% During the offline phase, \dhist{} processes the database schema with primary/foreign key information and optionally the query workloads to align and bin join keys, build single-table statistics, and find \topk{} join paths.
During the offline phase, \dhist{} processes the database schema — utilizing primary and foreign key information — and optionally the query workload to align and bin join keys, construct single-table statistics, and identify dominant join paths.
In the online inference phase, SQL queries are decomposed into tables and their corresponding attribute information. Using the table information, the single-table histograms are joined, with necessary filters applied for efficient cardinality estimation.
% In the online inference phase, SQL queries are processed by decomposing them into single-table statistics, retrieving and joining relevant histograms, and applying necessary filters for efficient AQP estimation.

\subsubsection*{\textbf{Offline State Building Phase}} 
During the state-building phase, primary/foreign key relationships in tables are analyzed, and popular join patterns from historical query workloads are optionally incorporated.
The bin structure for the same join keys across tables is aligned to ensure consistency, minimizing information loss during cross-bin interpolation and simplifying prediction complexity.
This study employs an equi-width binning strategy, with non-uniformity information captured via the \topk{} container (see \Cref{sec:data:structure} for details).

% For each table, \dhist{} histograms are constructed for all primary/join keys and, optionally, for relevant predicate attributes in historical query workloads. 
% For each table, \dhist{} histograms are constructed for all primary/join keys and relevant predicate attributes in historical query workloads. 
% Specifically, 1D \dhist{} histograms are built for primary/foreign keys, and 2D \dhist{} histograms are created for pairwise combinations of keys and non-key attributes within the same table.
% \textcolor{red}{
% \dhist{} will be constructed on all tables.
% Specifically, 1D \dhist{} histograms are built for all primary/join keys and relevant predicate attributes in historical query workloads, and 2D \dhist{} histograms are created for pairwise combinations of keys in join path and predicate attributes within the same table.
% }
\dhist{} will be constructed for all tables.
Specifically, 1D \dhist{} histograms are built for all primary/join keys, as well as for predicate attributes referenced in historical query workloads. 2D \dhist{} histograms are created for pairwise combinations of keys that appear in dominant join paths and predicate attributes within the same table.
% \textcolor{red}{delete something here???}
% We observe that improper handling of categorical attributes leads to significant errors, as filters on these attributes are typically highly selective, and the uniform assumption fails to produce accurate results.
Improper handling of categorical attributes causes significant errors, as filters on these attributes are highly selective, and the uniform assumption yields inaccurate results.
For categorical attributes, histograms are binned by all distinct values. 

This approach mirrors the frequency histograms \cite{oracleHistogram} used in Oracle databases but retains all values rather than only popular ones.
% \textcolor{red}{delete}
In this study, attributes are automatically classified as categorical if their cardinality falls below a specified threshold, though manual designation is also supported.

% The final step identifies \topk{} join keys within join paths – critical items whose mishandling would cause severe overestimation errors, as these items disproportionately influence join outcomes.
The final step identifies \topk{} join keys within join paths – critical items whose mishandling would cause severe overestimation errors.
% For implementation details, see \Cref{sec:dominating:join:key:discovery}.

\begin{table}
    \caption{Notation.}
    \footnotesize %scriptsize
    \label{tab:notation}
    \begin{tabularx}{\linewidth}{@{}p{0.12\textwidth}X@{}}

\toprule
  \underline{Notation:} \\
  Pure Cardinality & The cardinality of queries without filter predicates. \\
  Join Path & A set of primary/foreign keys that satisfy the join conditions. For example, $T_1.k_1 = T_2.k_2 = 1$ is a join path of query in \Cref{fig:workflow}. \\
  \underline{Parameters:} \\
  $k$ & The number of dominating items per bin. \\
  n   &  The number of histogram bins.\\
  NDV     & The \textbf{N}umber of \textbf{D}istinct \textbf{V}alues in a column. For example, if a column only contains the values 2, 2, 3, and 5, then the NDV for this column is 3. \\
  NV      & The \textbf{N}umber of \textbf{V}alues in a column. For example, if a column only contains the values 2, 2, 3, and 5, then the NV for this column is 4. \\
  BAC     &  The \textbf{B}ackground \textbf{A}verage \textbf{C}ounter, or the average number of values for the non-dominating items. For example, if a column only contains the values 2, 2, 2, 2, 3, 3 and 5, where 2 is the dominating item, 3, 3 and 5 are the background items, BAC is $(1+2)/2=1.5$.\\
  MFV     &  The \textbf{M}ost \textbf{F}requent \textbf{V}alue.\\
  % background &\\
\bottomrule
    \end{tabularx}
    % \vspace{-0.5cm}
\end{table}

% \vspace{-0.2cm}
\subsubsection*{\textbf{Online Inference Phase}}
When a query arrives, \dhist{} parses and decomposes it into individual component tables.
% Next, relevant single-table join key \dhist{} histograms are joined. 
% \textcolor{red}{
% Next, the relevant histograms of join key \dhist{} are joined. 
% Filter predicates are then applied to calibrate the dominant join paths and calculate the selectivity.
% }
Next, the relevant \dhist{} histograms corresponding to join keys are joined. Filter predicates are subsequently applied to calibrate the dominant join paths and compute selectivity.
% \textcolor{red}{
% We compute the cardinality estimation result by multiplying the size of the joined tuples by the selectivity.
% Details on the selectivity calculation can be found in \Cref{sec:dim:reduction}.
% }
We compute the cardinality estimation result by multiplying the size of the joined tuples by the selectivity.
Details on the selectivity calculation can be found in \Cref{sec:dim:reduction}.
The filter calibration stage is essential; omitting it causes catastrophic overestimation, especially when filter predicates strongly correlate with dominating join paths. 
This error pattern is quantitatively analyzed in \Cref{fig:correlation:effect}.

\vspace{-0.2cm}
\subsection{Dimension Reduction of Single Table }
\label{sec:dim:reduction}
In this paper, we relax the attribute independence assumption with a conditional independence assumption relative to primary/join keys (PK/JK), which fits in well with the actual table distribution. 
% During the online inference phase, instead of learning the accurate high-dimensional attribute distribution,
% \dhist{} relies on 1D or 2D histograms to construct an approximation of the actual distribution.
% We build 1D \dhist{} for all primary/foreign keys, and 2D \dhist{} for all pairwise combinations between primary/foreign keys and non-key attributes.
% \textcolor{red}{....}
Take $T_2$ in \Cref{fig:workflow} as an example. 
Instead of constructing the high-dimensional data distribution of $T_2$, only the smallest relevant set of \dhist{} histograms is used.
In this example, we use 1D \dhist{} to get $P(k_2)$ and 2D \dhist{} to get $P(k_2,y)$ and $P(k_2,z)$, and then use the following equation to calculate the probability $P(k_2,y,z)$:
% In this example, we need to construct the probability $P(k_2,y,z)$ by 1D \dhist{} $P(k_2)$ and 2D \dhist{} $P(k_2,y)$, $P(k_2,z)$, by the following equation:
% \begin{align}
% \begin{split}
%     P(k_2,y,z)&=P(y,z|k_2)\cdot P(k_2)\\
%     &\approx P(y|k_2)\cdot P(z|k_2)\cdot P(k)\\
%     &=\frac{P(y,k_2)}{P(k_2)}\cdot \frac{P(z,k_2)}{P(k_2)}\cdot P(k_2)\\
%     &=\frac{P(y,k_2)\cdot P(z,k_2)}{P(k_2)}
% \end{split}
% \end{align}
\begin{align}
\begin{split}
    P(k_2,y,z)&=P(y,z|k_2)\cdot P(k_2)\\
    &\approx P(y|k_2)\cdot P(z|k_2)\cdot P(k_2)\\
    &=\frac{P(y,k_2)\cdot P(z,k_2)}{P(k_2)}
\end{split}
\end{align}
where $k_2$ is the PK/JK in the query, y and z correspond to filter predicates involving attributes y and z. In this way, high-dimensional data distributions are factorized into the combination of 1-2D \dhist{}s, with the following assumption:
% \begin{assumption}[Conditional Independence of Non-key Attributes]
% Given a relation $R$, the conditional distributions of non-key attributes with respect to 
% primary keys (PK) or joint keys (JK) are mutually independent. That is, for any non-key 
% attributes $A_1, A_2, \ldots, A_n$ and key $K$ (either PK or JK):
% $$P(A_1, A_2, \ldots, A_n \mid K) = \prod_{i=1}^n P(A_i \mid K)$$
% \end{assumption}
\begin{assumption}[Conditional Independence of Non-key Attributes]
Given a relation $R$, the conditional distributions of non-key attributes with respect to 
primary keys (PK) or join keys (JK) are mutually independent. That is, for any non-key 
attributes $A_1, A_2, \ldots, A_n$ and key $K$ (either PK or JK):
$$P(A_1, A_2, \ldots, A_n \mid K) = \prod_{i=1}^n P(A_i \mid K)$$
\end{assumption}

This assumption is stricter than the commonly adopted attribute dependence assumption, as it requires that PK/JK fully capture all correlations between non-key attributes. In other words, any dependence between attributes can be mediated through their relationship with the key attributes.

This dimension reduction mechanism can be easily applied to cases with multiple non-key attributes $y, z, p, q$, which is  
\begin{equation}
    P(k_2,y,z,p,q) \approx \frac{P(y,k_2)\cdot P(z,k_2)\cdot P(p,k_2)\cdot P(q,k_2)}{P^3(k_2)}
\end{equation}
% Here we relax the attribute independence assumption with a conditional independence assumption relative to keys, which fits in well with the actual table distribution. 
Specifically, $P(y,z|k_2)$ is approximated as $P(y|k_2)\cdot P(z|k_2)$, where $k_2$ is the primary/foreign key in $T_2$.

\subsection{\dhist{} Data Structure}
\label{sec:data:structure}
\begin{figure}[ht!]
    \centering
    \includegraphics[width=\linewidth]{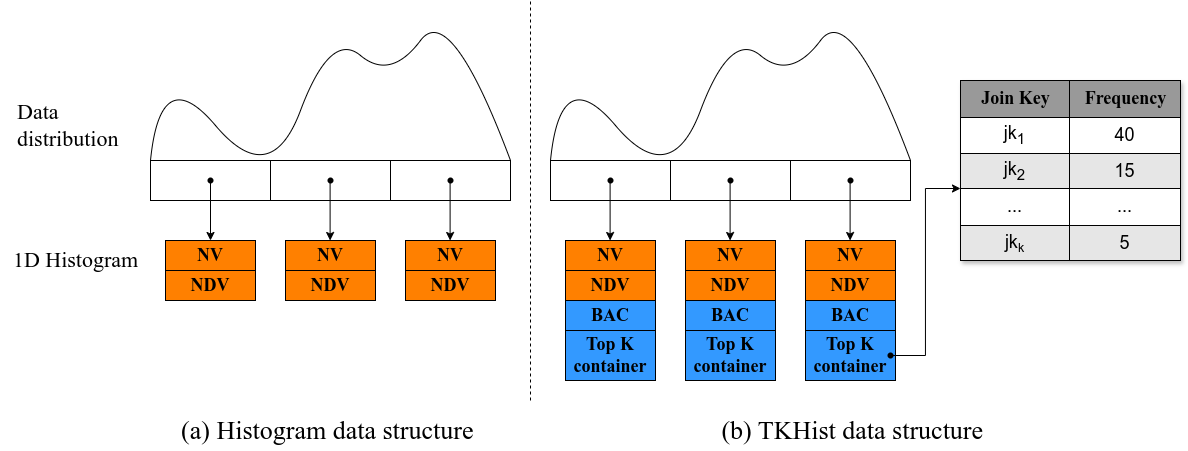}
    \caption{\dhist{} data structure vs histogram data structure.}
    \label{fig:structure}
    \vspace{-0.2cm}
\end{figure}

\Cref{fig:structure} compares the data structure of a \dhist{} with an equi-width histogram,  both containing $n$ bins.
% for a given data distribution. 
For an equi-width histogram, the space complexity is $\mathcal{O}(n)$, as illustrated in \Cref{fig:structure}.(a).

% As shown in \Cref{fig:structure}.(b), the data structure of \dhist{} resembles equi-width histograms but includes an additional container to track top-$k$ frequent (dominant) join keys, which contribute to the \topk{} join paths in the final results. 
As shown in \Cref{fig:structure}.(b), the data structure of \dhist{} includes an additional container to track top-$k$ frequent join keys, which contribute to the dominant join paths in the final results. 
This container uses a hash structure with $\mathcal{O}(1)$ time complexity for lookups/insertions, and $\mathcal{O}(k)$ space per bin, resulting in an overall $\mathcal{O}(nk)$ space overhead. 
% \textcolor{red}{delete}
% Experiments suggest $k$ typically ranges between 10--20, making the space increase manageable.
Notably, NV and NDV exclude \topk{} keys in \dhist{}.
% BAC(Background Average Counter) represents the average frequency of distinct values excluding \topk{} keys.
BAC(Background Average Counter) represents the average frequency of distinct values excluding \topk{} keys, which is crucial for multi-table query estimation.
% As will be demonstrated in the experimental part, BAC is crucial for accurate multi-table query estimation. Otherwise, some \topk{} join paths might be missed, causing significant errors. 
% The 2D \dhist{} structure follows similarly to the 1D case and is omitted for space reasons.
In this paper, we use equi-width histogram binning strategy for join tasks.
% For join queries, consistent bin boundaries across tables are necessary to avoid information loss during cross-bin interpolation. 
% While DBMSs often use complex binning strategies, including MHist \cite{deshpande2001independence,bruno2001stholes} and GBSA\cite{wu2023factorjoin}, to minimize intra-bin variance, we show that equi-width binning suffices when augmented with \topk{} containers to capture intra-bin non-uniformity - a conclusion validated empirically.

\subsection{Join Query Processing}
\label{sec:join:path:processing}
\begin{figure}[h!]
    \vspace{-0.4cm}
    \centering
    \includegraphics[width=\linewidth]{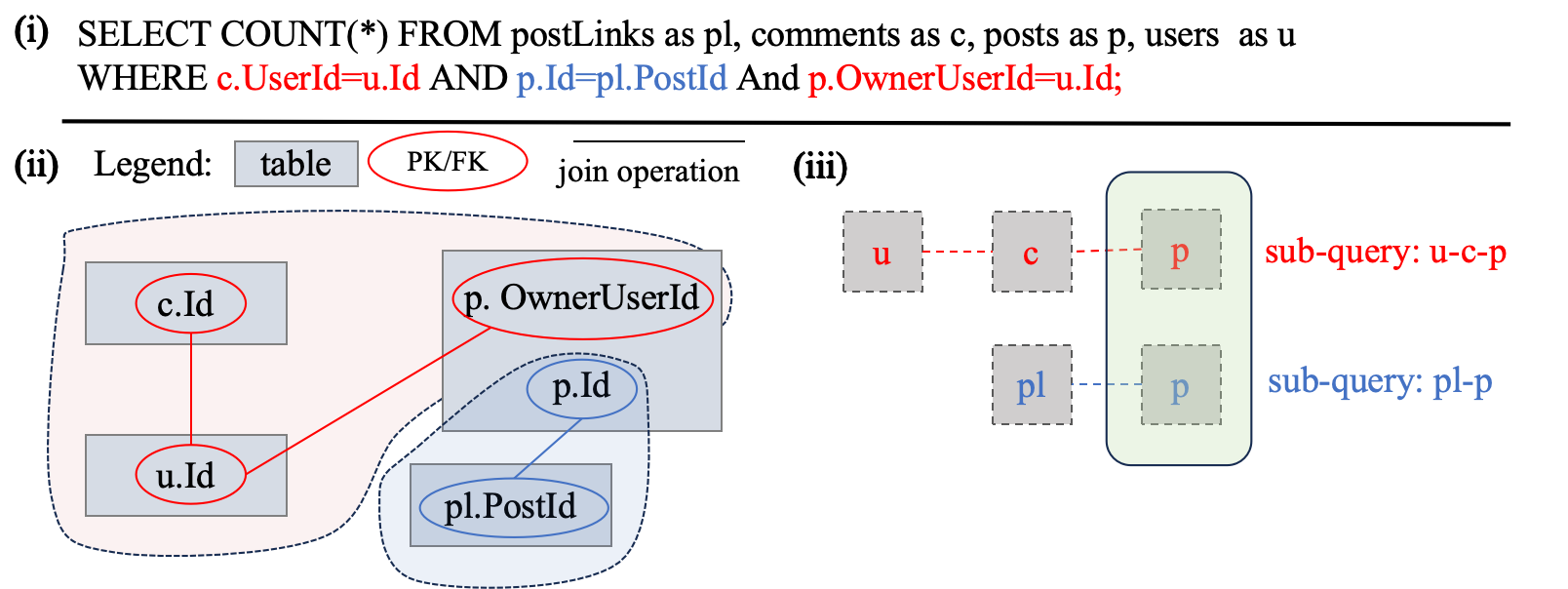}
    
    \caption{Decomposition of multi-table join queries into sub-queries.}
    \label{fig:join:path}
    \vspace{-0.2cm}
\end{figure}

Multi-table join queries typically involve complex combinations of filter predicates, inner/outer joins, and both cyclic/acyclic patterns. 
This paper focuses on acyclic equi-joins combining chain and star join patterns. 
On the online inference phase, our approach decomposes queries into sub-queries along shared join attributes for efficient processing.

\Cref{fig:join:path}.(i) illustrates a typical SQL query from the STATS-CEB workload that joins 4 tables.
\Cref{fig:join:path}.(ii) visualizes the join as a graph where the color-matched tables share the same join keys.
For instance, table u, c and p form a star join on join key u.Id.
\Cref{fig:join:path}.(iii) shows the decomposed sub-queries: u-c-p and pl-p.
These sub-queries are finally chained together by two join keys (p.Id and p.OwnerUserId) in table p.

\section{Implementation and Optimization}
\label{sec:impl}
% \textcolor{red}{github anonymous}
The source code of \dhist{} is publicly available \footnote{\textcolor{black} {\url{https://github.com/qingzma/dhist}}}, along with several datasets and pre-trained models.
% This section discusses the efficient implementation and optimization of \dhist{} during its online inference phase. 
% Several challenges must be addressed, including query processing of \dhist{} for two-table join and its extension to multi-table join, relaxation of correlation between join keys and filter predicates, \dhist's hyper-parameters tuning,  update support, etc.
% In this section, we aim to discuss the efficient implementation and optimization of \dhist{}, including \dhist{} join algorithm, discovery of correlation between dominant join keys and filter predicates, update support, etc.
In this section, we aim to discuss the efficient implementation and optimization of \dhist{}, including \dhist{} join algorithm, relaxation of correlation between join keys and filter predicates, update support, etc.

\subsection{Join of \dhist}
\label{sec:join:dhist}
Join-Histogram often yields significant errors when applied to skewed distributions.
In this study, \dhist{} addresses this issue by relaxing the uniformity assumption for join keys within each bin and introducing a \topk{} container (see \Cref{fig:structure}).

% Similarly to the Join-Histogram, which uses the Selinger method for bin-wise predictions,
% this study adopts the same strategy for background (non \topk{}) items, which is 
% % \begin{equation}
% %     Selinger(Ta \bowtie Tb) = \frac{|Ta|\cdot |Tb|}{max(NDV(Ta), NDV(Tb))}
% % \end{equation}
% \textcolor{red}{change}
% \begin{equation}
%     Selinger(Ta \bowtie Tb) = \frac{|Ta|\cdot |Tb|}{max(NDV(Ta), NDV(Tb))}
% \end{equation}
% where NDV is the number of distinct values (see \Cref{tab:notation}).
 
This study adopts Selinger method to calculate BAC value, which is 
% \begin{equation}
%     Selinger(Ta \bowtie Tb) = \frac{|Ta|\cdot |Tb|}{max(NDV(Ta), NDV(Tb))}
% \end{equation}
% \textcolor{red}{change}
% \begin{equation}
%     BAC = Selinger(Ta \bowtie Tb) = \frac{|Ta|\cdot |Tb|}{max(NDV(Ta), NDV(Tb))}
% \end{equation}
\begin{equation}
     \frac{|Ta|\cdot |Tb|}{max(NDV(Ta), NDV(Tb))}
\end{equation}
% where NDV is the number of distinct values (see \Cref{tab:notation}).
where NDV is the number of distinct values exclude keys in \topk{} container (see \Cref{tab:notation}).

For \topk{} join keys, we aim to produce accurate cardinality calculation, while an over-estimation strategy is applied for robust prediction.
% For \topk{} join keys that do not appear in other histograms, the BAC is used to correct errors, ensuring that popular items in one table still have the opportunity to join, even if they are sparse in another table.
% The complete process is summarized in the JTKH Algorithm below:
For \topk{} join keys that do not appear in other histograms, the BAC is used to correct errors, ensuring that popular items in one table still have the opportunity to join.
The complete process is summarized in the JTKH Algorithm below:

\begin{algorithm}
\footnotesize
\caption{Join \dhist{} Algorithm (JTKH)}
\label{alg:join:dhist}
\begin{algorithmic}[1]
\Require  $TKHist_a$ and $TKHist_b$ to be joined, with the corresponding top-$k$ join key containers $c_{a,i}$, $c_{b,i}$, 
background average counter (BAC)
within $bin_{a,i}$, $bin_{b,i}$ .
\Ensure The collection of estimated join cardinality per bin.
\State counter $\gets$ []
\For{($bin_{a,i}$, $bin_{b,i}$) pair in ($TKHist_a$, $TKHist_b$)}
\State bin\_counter $\gets$ 0
\item[]

\LeftComment{For background values, perform Selinger estimation} 
\State  bin\_counter += Selinger($bin_{a,i}$, $bin_{b,i}$)
\item[]

\LeftComment{For dominating values, perform key-value multiplication}
\For{ common join key $k$ in container $c_{a,i}$ and  $c_{b,i}$}
\State bin\_counter += $c_{a,i}[k] \times  c_{b,i}[k]$ 
\EndFor

\For{  join key $k$ only in container $c_{a,i}$ }
\State bin\_counter += $c_{a,i}[k] \times  BAC_{b,i}$ 
\EndFor

\For{  join key $k$ only in container $c_{b,i}$ }
\State bin\_counter += $c_{b,i}[k] \times  BAC_{a,i}$ 
\EndFor
\State counter.append(bin\_counter)
\EndFor
\State return counter
\end{algorithmic}
\end{algorithm}

By using binary join framework, the JTKH Algorithm can be naturally extended to multi-table joins.
For example, as the sub query $u-c-p$
shown in \Cref{fig:join:path}, we first compute the intermediate histogram ${\dhist}(u-c)$.
The intermediate result is then joined with table $p$ to generate the final composite histogram ${\dhist}(u-c-p)$.
The NDV for the intermediate \dhist{} of is updated using the Selinger model:
\begin{equation}
    NDV(Ta \bowtie Tb)=min(NDV(Ta), NDV(Tb))
\end{equation}

Meanwhile, the result of \Cref{alg:join:dhist} can be regarded as the dominant join paths, which plays a crucial role in processing the filter predicates in our approach.
Traditional aggregation of join paths is infeasible due to exponential join size growth. 
For example, a five-table join query in the STATS-CEB workload generates up to $3\times 10^{12}$ tuples,  requiring hours to execute on a machine with 128GiB RAM, 
while six-table joins cause memory errors in Postgres \cite{stonebraker1990implementation} and MonetDB \cite{nes2012monetdb},
while \dhist{} with \Cref{alg:join:dhist} can compute five-table STATS-CEB joins in 25ms and six-table joins in 30ms.

\vspace{-0.2cm}
\subsection{Discovery of Correlation between Dominant Join Path and Filter Predicates}
\label{sec:dominating:join:key:discovery}
As discussed in \Cref{fig:top:k}, top-10 join paths account for 2.9\% of final join results in two-table joins, 
but this proportion rises dramatically to 76\% for three-table joins, a critical factor in the join results. 
Failure to properly handle filter predicates correlated with these dominant join paths leads to significant over-estimation.
% Tracking these \topk{} join paths is therefore essential for queries involving related filters.
% Hence, tracking these dominant join paths is essential for queries involving related filters.
Hence, tracking the correlation between these dominant join paths and filter predates is essential for estimation result.

Moreover, we only maintain the longest join query template is considered.
For instance, in the STATS-CEB workload, we only maintain the attributes of dominant join paths for the longest join query $u-b-c-v-p-ph$, as it covers other sub-queries like $u-c$, $u-c-v$, etc.
\begin{algorithm}[th!]
\footnotesize
% \caption{Top-$k$ Join Path Discovery Algorithm (TJPD)}
% \caption{Dominant Join Path Discovery Algorithm (DJPD)}
\caption{Dominant Join Path Correlation Discovery Algorithm (DJPCD)}
\label{alg:dominating}
\begin{algorithmic}[1]
\Require  \dhist{} for all tables in a dataset, with the longest join query templates.
% \Ensure corrected cardinality estimation excluding irrelevant top-$k$ join paths% $counter$
\Ensure Corrected TKHist excluding irrelevant top-$k$ join keys% $counter$
\item[] 
\LeftCommentNew{Offline phase, find correlations between dominant join path and filter predicates} 
% \State hists $\gets$ dict()
\For{query in longest join query templates}
\State hist=JTKH(query)
\State dominant join path $\gets$ hist.getTopJoinKeys()
\For{table $t_i$ in query}
\State correlations $\gets$ dict()
\For{any possible filter attribute $attr_j$ in table $t_i$}
\State df $\gets$ table $t_i$'s relevant tuples that contain join keys in dominant join path
\State correlation $\gets$ df.buildDictBetweenJkAndAttribute()
\State table.correlations[$jk$][$attr_j$]=correlation
\EndFor
\EndFor
\EndFor
\item[]
\LeftCommentNew{Online Query Processing}
% \item[] 
\LeftCommentNew{Step 1: Discover irrelevant top join keys}
\State TopJKsExcluded $\gets$ []
\For{(table, attr, filter) in a new query}
\State    correlation $\gets$ table.correlations[$jk$][$attr_j$]
\State    JksExcluded = correlation.notSatisfy(filter)
\State    TopJKsExcluded.append(JksExcluded)
\EndFor
% \item[]
\LeftCommentNew{Step 2: Exclude irrelevant top join keys}
\State modify JTKH algorithm to exclude TopJKsExcluded
\item[]
\State return corrected estimation
\end{algorithmic}
\end{algorithm}

% \dhist{} enables efficient dominant join path discovery, computing five-table STATS-CEB joins in 25ms and six-table joins in 30ms.
% Moreover, we do not maintain top-$k$ join paths for all possible combinations in the query workloads.
% Instead, only the longest join query template is considered.
% For instance, in the STATS-CEB workload, we only maintain the attributes of top-$k$ join paths for the longest join query $u-b-c-v-p-ph$, as it covers other sub-queries like $u-c$, $u-c-v$, etc.
% Moreover, we only maintain the longest join query template is considered.
% For instance, in the STATS-CEB workload, we only maintain the attributes of dominant join paths for the longest join query $u-b-c-v-p-ph$, as it covers other sub-queries like $u-c$, $u-c-v$, etc.

\Cref{alg:dominating} outlines the steps to discover the correlation between dominant join paths and filter predicates.
In the offline phase, the relationship between join keys and filter predicates for each table is discovered.
For a foreign key with multiple occurrences within a table, we select the minimum attribute value instead of the average or maximum values as the former tends to over-estimates the cardinality, which potentially produces a more robust query execution plan in other application scenario.
During the online query processing phase, join paths that do not satisfy the filter predicates are identified and excluded during JTKH inference. 
% We adopt a strategy to exclude irrelevant join keys instead of including relevant join keys as this potentially produces a more robust over-estimation.

% \subsection{Hyper-parameters Tuning}
% \dhist{} has two main hyper-parameters: the number of bins and the $k$ value in the top-$k$ container, which balance accuracy and space overhead.
% Increasing the number of bins reduces prediction error, at the cost of linearly growing model size and query latency. 

% \textcolor{red}{delete???}
% The selection of $k$ is worth further investigation.
% When $k$ = 0, \dhist{} becomes normal histograms and our method is degraded to the well-known Join-Histogram method.
% When $k$ is unbounded, our method provides accurate estimates using exact frequency data.
% In other words, as $k$ increases, our method asymptotically converges to an unbiased estimate.
% In our experiments(see \Cref{sec:experiment}), a bin size of 100-200 and a $k$ value of 10-20 yield accurate estimates.

\subsection{Support for updates}
\label{sec:update:discussion}
\dhist{} efficiently handles frequent updates by maintaining single-table statistics.
Using an equi-width binning strategy, locating the target bin requires $\mathcal{O}(1)$ time through linear interpolation.
For tuples containing dominating join keys, counter updates take $\mathcal{O}(1)$ time via hash table operations.
% If the new tuple is insignificant and does not belong to \topk{} keys, updating BAC takes $\mathcal{O}(1)$ time.
If the new tuple does not belong to \topk{} keys, updating the BAC value takes $\mathcal{O}(1)$ time.
Overall, tuple updates have $\mathcal{O}(1)$ time complexity.

A current limitation exists for emerging dominant keys: newly frequent join keys introduced during incremental updates remain classified as background values. 
We recommend periodic \dhist{} rebuilds to capture such distribution shifts.

\subsection{Error Guarantee}
% \textcolor{red}{....}
As will be demonstrated in \Cref{sec:tuning}, \dhist{} enables space-accuracy trade-offs through hyper-parameter tuning. 
In addition, a probabilistic error bound $\epsilon$ could be provided.
Take the 1D \dhist{} structure in \Cref{fig:structure} as an example.
The exact cardinality for each bin can be expressed as:
% \begin{equation}
%     |bin_i|=BAC_i\cdot NDV_i + \sum_{k\ in\ top-k\ container}{Frequency(k)}
% \end{equation}
\begin{equation}
    |bin_i|=BAC_i\cdot NDV_i + \sum_{k\, in\, top-k\, container_i}{Frequency(k)}
\end{equation}
where the second term in RHS is computed exactly during join processing, and the first term uses the Selinger model approximation.
To ensure an overall error bound $\epsilon$, we constrain the approximation error per bin $\epsilon_i$ such that:
\begin{equation}
    \frac{BAC_i\cdot NDV_i}{|bin_i|}<\epsilon_i
    \label{equ:bin}
\end{equation}

Assuming statistical independence between bin errors, the total error bound follows the Root Sum of Squares (RSS) principle: 
\begin{equation}
\label{equ:rss}
    \epsilon=\sqrt{n}\cdot \epsilon_i
\end{equation}
where n is the number of bins. 
For worst-case analysis, $\epsilon= n\cdot \epsilon_i$.
Combining \Cref{equ:bin,equ:rss} yields
\begin{equation}
    \sqrt{n}\cdot \frac{BAC_i\cdot NDV_i}{|bin_i|}<\epsilon
\end{equation}
This statistical error guarantee applies specifically to single-table queries, with slightly increased space requirements. 
Join queries lack formal error guarantees due to complex correlation effects. 
However, experimental results demonstrate that \dhist{} provides accurate estimates for pure joins and outperforms state-of-the-art methods.
\section{Experiments}
\label{sec:experiment}

\subsection{Experimental Setup}
\label{sec:setup}
% \subsubsection*{\textbf{Datasets}} 
\subsubsection*{\textbf{Benchmarks}} 
We evaluate \dhist's performance over two well-known benchmarks: STATS-CEB \cite{han2021cardinality} and IMDB/JOB-light \cite{kipf2018learned}.
% \textcolor{red}{add a table here}

The STATS-CEB benchmark consists of 8 real-world tables from Stack Overflow and related websites, featuring 146 star \& chain queries. 
These tables exhibit skewed distributions and highly correlated attributes, containing up to millions of tuples. The resulting join outputs range from 200 to $2 \cdot 10^{10}$ tuples. 
In addition, the pure cardinality (join queries without filter predicates) can be up to $10^{15}$, which is a challenging task.

The IMDB/JOB-light benchmark consists of six tables extracted from the real-world IMDB dataset, featuring 70 star-schema queries.
The resulting join outputs range from 9 to $9 \cdot 10^{9}$.

\subsubsection*{\textbf{Baselines}} 
% We compare the performance of \dhist{} against SOTA cardinality estimation methods, including:
We compare the performance of \dhist{} against SOTA cardinality estimation methods, including:
\begin{enumerate}[noitemsep,wide=0pt, leftmargin=\dimexpr\labelwidth + 2\labelsep\relax, label={\arabic*)}]
    \item TrueCard refers to the true cardinality without error, serving as the ground truth for comparison.
    % \item PostgreSQL \cite{stonebraker1990implementation} uses the histogram-based cardinality estimation based on uniformity and independency assumptions(referred to as Postgres in the rest of this paper).
    \item PostgreSQL \cite{stonebraker1990implementation} uses the histogram-based cardinality estimation based on uniformity and independence assumptions.
    \item Oracle\cite{oracle23ai} uses several types of histograms: frequency, top frequency, height-balanced, and hybrid. The histograms will be created based on the distribution of data.
    % \item Join-Histogram \cite{adellera} is the well-known method to join histograms and relax the join key uniformity assumption. %, and is \textcolor{red}{adopted in the Oracle database}.
    % \item DeepDB\cite{hilprecht2020deepdb}, FLAT\cite{zhu2021flat} are learned data-driven methods that use fanout techniques to pass join key distributions of various query templates.
    \item WJSample\cite{zhao2018random} uses the wander join sampling techniques to generate uniform samples of the join results. 
    \item DeepDB\cite{hilprecht2020deepdb} and BayesCard \cite{wu2020bayescard} are learned data-driven methods that use fanout techniques to pass join key distributions of various query templates.
    % \item MSCN\cite{kipf2018learned} is the query-driven method that learns the cardinality of queries based on history query workloads.
    \item FactorJoin\cite{wu2023factorjoin} uses single-table statistics to estimate an upper bound for true cardinality.

\end{enumerate}

% \textcolor{red}{Notably, the PostgreSQL, Oracle and WJSample mentioned above are deployed through Docker.}

\subsubsection*{\textbf{Metrics}}
We evaluate \dhist{} using two complementary metrics:
\begin{enumerate}[noitemsep,wide=0pt, leftmargin=\dimexpr\labelwidth + 2\labelsep\relax, label={\arabic*)}]
\item Q-error \cite{moerkotte2009preventing}: A widely adopted metric for cardinality estimation, defined as:
\begin{equation}
    Q{\text -}error = \frac{\max(estimate, true)}{\min(estimate, true)}
\end{equation}
By definition, Q-error is always greater or equal to 1. 
While effective for measuring error magnitude, this metric loses directional information about under/over-estimation
\item $ratio$: To explicitly quantify estimation directionality, we use the accuracy ratio:
\begin{equation}
    ratio=\frac{estimate}{true}
\end{equation}
where $ratio > 1$ indicates overestimation, and vice versa.
\end{enumerate}

\subsubsection*{\textbf{Environment}}
% All experiments are run on a Fedora Server with AMD Ryzen 9 7950X CPU with 16 cores and 32 threads, 128GB DDR5 memory, and 2TB SSD. 
% \dhist{} is implemented in single-threaded Python. 
% Numpy and other packages are used to optimize the query processing phase.
% We use open-source implementation of all competitors and tune their parameters to achieve the best possible performance.
% During the end-to-end benchmark evaluation, we use a modified version of PostgreSQL that allows cardinality estimation results to be injected into database systems to optimize query plans \cite{han2021cardinality}. 
% All queries are executed three times and we report the average performance.
All experiments were conducted on a Fedora Server equipped with an AMD Ryzen 9 7950X CPU (16 cores, 32 threads), 128 GB DDR5 memory, and a 2 TB SSD.
\dhist{} is implemented in single-threaded Python. 
% Numpy and other packages are utilized to optimize the query processing phase.
We use open-source implementation of all competitors and tune their parameters to achieve the best possible performance.
For end-to-end benchmark evaluation, a modified version of PostgreSQL is used, enabling cardinality estimation results to be injected into the database system to optimize query plans \cite{han2021cardinality}.
All queries are executed three times, and we report the average performance.

\subsection{Overall performance}
% \subsection{\textbf{End-to-end performance}}
\label{sec:end:to:end}
% We follow the methodology of a recent work \cite{han2021cardinality} to perform an end-to-end comparison of \dhist{} with state-of-the-art competitors.
\subsubsection*{\textbf{End-to-end performance}}
We follow the methodology of a recent work \cite{han2021cardinality} to perform an end-to-end comparison of \dhist{} with baselines.
% TrueCard (the ground truth) represents the optimal query plan and end-to-end time baseline.
% Query execution time and planning time are reported separately for clarity.
End-to-end time is calculated as the sum of execution time and planning (including model inference time).
Since sub-query predictions are precomputed and injected into the database core, PostgreSQL's planning time excludes the time required for different methods to process the sub-queries.
% We manually incorporate the overall prediction time into the planning time for a fair comparison.

% \Cref{tab:end_to_end} summarizes the end-to-end performance across methods for both STATS-CEB and IMDB/JOB-light datasets.
\Cref{tab:end_to_end} summarizes the end-to-end performance of baselines on STATS-CEB and IMDB/JOB-light.
In both benchmarks, the performance of \dhist{} and FactorJoin closely approximates TrueCard, whereas other methods require more time to execute.

\begin{table}[!htbp]
    \caption{End-to-end performance on STATS-CEB and IMDB/JOB-light benchmarks.}
        \footnotesize 
        \label{tab:end_to_end}

    \centering
    \scalebox{1.0}{\begin{tabular}{c|c|ccc}
    \toprule
    % \hline
    Benchmark & Method & End-to-End(s) & Execution(s) & Planning(s) \\
    \hline
     \multirow{8}*{STATS-CEB}
    
    & TrueCard         & 7826 & 7826 & 0.2   \\
    & Postgres      & 11584 & 11581 & 2.84    \\
    & Oracle        & 14580 & 14567 & 12.44    \\
    & WJSample      & 15553 & 15481 & 72.15    \\

    & DeepDB        & 9208 & 9198 & 9.29    \\
    & BayesCard     & 11487 & 11479 & 8.25    \\
    & Factorjoin    & 8319 & 8308 & 11.16    \\
    & TKHist(Ours)   & \textbf{8212} & 8186 & 26    \\
    \hline
    \multirow{8}*{IMDB/JOB-light}
    
    & TrueCard         &  2438& 2438 & 0.04 \\
    & Postgres      &  2662& 2662 & 0.05  \\
    & Oracle        &  2648& 2647 & 1.28 \\
    & WJSample      & 2708 & 2687 & 21.12   \\ 
    & DeepDB        &  2634& 2634 & 0.64 \\
    & BayesCard     &  2576& 2575 & 0.96   \\    
    & Factorjoin    &  2559& 2558 & 1.08 \\
    & TKHist(Ours)   & {\textbf{2481}} & 2474 & 6.97 \\

    \bottomrule
    \end{tabular}}
    \vspace{-0.2cm}
\end{table}

Moreover, a small subset of queries in STATS-CEB dominates the overall execution time \cite{kim2024asm}. 
Optimizing these critical queries is crucial for peak performance, but their dominance may reduce the benchmark’s ability to differentiate method effectiveness.
To address this, for the STATS-CEB benchmark, a threshold of 3000 seconds is applied to each query's execution time.
If a query times out, this threshold is used to compensate for the execution time.
% In both benchmarks, the performance of \dhist{} and FactorJoin closely approximates TrueCard, whereas other methods require more time to execute.

\begin{figure}[!htbp]
    \centering
    \includegraphics[width=0.8\linewidth]{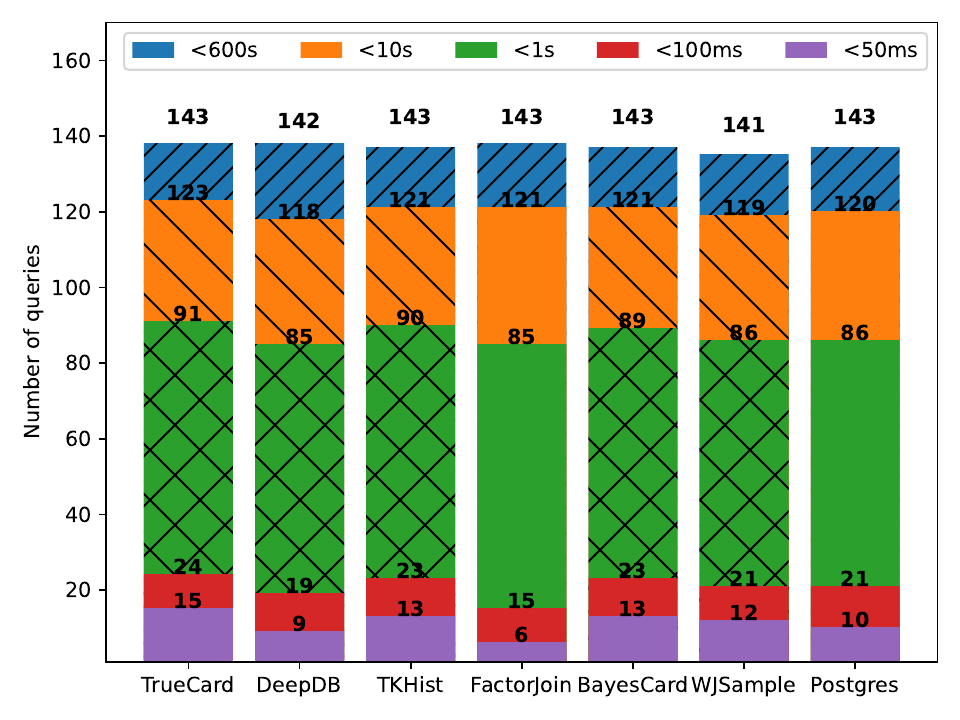}
    % \caption{End-to-end query latency distribution for STATS-CEB.}
    \caption{Distribution of end-to-end execution times for queries in STATS-CEB workload.}
    \label{fig:end:to:end:stats:time:split}
    % \vspace{-0.4cm}
\end{figure}

\begin{table*}[h]
% \begin{table}[!htbp]
    % \caption{Q-errors for sub-queries in STATS-CEB and JOB-light Benchmark}
    \caption{Overall performance of sub-queries in the STATS-CEB and JOB-light benchmarks.}
        \footnotesize %scriptsize
        \label{tab:qerror}
    \centering
    % \begin{tabular}{c|c|c|c|c|c|c}
    \scalebox{1.1}{
    % \scalebox{0.7}{
    \begin{tabular}{c|c|ccccc|c|c}
    \toprule
    % \hline
    % Benchmark & Method & median & 90th & 95th & 99th & Max & \color{red}{Latency(ms)} & Model Size(MB)\\
    Benchmark & Method & median & 90th & 95th & 99th & Max & Latency(ms)& Size(MB) \\
    \hline
    \multirow{7}*{STATS-CEB}
    % & DHist(Ours)   & 1.57 & \textbf{8.27} & \textbf{13.30} & \textbf{37.16} & \textbf{141.86} & - & - \\
    % & DHist(Ours)   & 1.67 & \textbf{7.09} & \textbf{10.53} & \textbf{28.43} & \textbf{81.02} & 、\color{red}{10.53} & 1.4MB \\
    % & DHist(Ours)   & 1.67 & {7.09} & {10.53} & \textbf{28.43} & \textbf{81.02} & 、\color{red}{10.53} & 1.4MB \\
    
    % & DHist(Ours)   & 1.67 & {7.09} & {10.53} & \textbf{28.43} & \textbf{81.02} & {10.53} & 2.7 & 21\\
    & Postgres      & 1.94 & 29.87 & 85.60 & 1317.16 & 7612121.49 & 1.09 & - \\
    & Oracle        & 3.36 & 148.38 & 610.03 & 8749.13 & 918205.47 & 4.78 & - \\
    % & WJSample      & 1.97 & 31.17 & 92.73 & 1293.54 & 6211584.29 & 0.86 & - \\
    & WJSample      & \textbf{1.09}& \underline{3.29} & \underline{9.08} & 384.56 & 644030 & 27.55 & - \\
    & DeepDB        & 1.60 & 10.13 & 28.36 & 150.46 & 22375.17 & 3.57 & 74.7 \\
    % & DeepDB        & 1.60 & 10.13 & 28.36 & 150.46 & 22375.17 & 3.57 & 74.7 & 1259\\
    % & FLAT          & 1.66 & 10.34 & 23.48 & 645.76 & 118883.23 & - & - \\
    % & BayesCard     & \textbf{1.18} & 44.95 & 92.19 & 154.17 & 897.10 & 3.17 & - \\
    % & BayesCard     & \textbf{1.20} & \textbf{2.91} & \textbf{6.46} & 64.27 & 839.09 & 3.17 & 5.95MB \\
    & BayesCard     & {\underline{1.20}} & \textbf{2.91} & \textbf{6.46} & \underline{66.56} & \underline{839.09} & 6.64 & 6.3 \\
     % & BayesCard     & {1.20} & \textbf{2.91} & \textbf{6.46} & 66.56 & 839.09 & 6.64 & 6.3 & 81\\
    % & Factorjoin    & 2.57 & 28.77 & 81.12 & 1031.74 & 980918.15 & 4.29 & 2.4MB \\
    & Factorjoin    & 2.42 & 26.67 & 71.95 & 869.03 & 980777.77 & 4.87 & 6.6 \\
    % & Factorjoin    & 2.42 & 26.67 & 71.95 & 869.03 & 980777.77 & 4.87 & 6.6 & 26\\
    & TKHist(Ours)   & 1.67 & {7.09} & {10.53} & \textbf{28.43} & \textbf{81.02} & {10.53} & 2.7 \\
    \hline
    \multirow{7}*{IMDB/JOB-light}
    % & DHist(Ours)   & 1.40 & 3.84 & 6.23 & 16.36 & 55.04 & \color{red}{10.95} & 787KB \\
    
    & Postgres      & 2.61 & 33.32 & 92.99 & 819.42 & 3511.50 & 0.89 & - \\
    & Oracle        & 2.86 & 24.33 & 102.89 & 871.22 & 18022.33 & 1.79 & - \\
    % & WJSample      & 2.28 & 21.06 & 44.23 & 701.52 & 3606.40 & 0.39 & - \\
    & WJSample      & \underline{1.16} & 72.0 & 3422 & 355667 & 1499619 & 20.37 & - \\
    & DeepDB        & 1.26 & \textbf{2.14} & \textbf{4.13} & 49.61 & 72.00 & 0.95 &  39.1\\
    % & FLAT          & 1.66 & 10.34 & 23.48 & 645.76 & 118883.23 & - & - \\
    % & BayesCard     & \textbf{1.14} & 2.84 & 4.82 & 10.33 & 42.51 & 1.33 & 23.3 \\
    & BayesCard     & \textbf{1.14} & \underline{2.84} & \underline{4.82} & \textbf{10.33} & \textbf{42.51} & 1.33 & 1.6 \\
    % & Factorjoin(sample)    & 11.02 & 338.36 & 949.58 & 14132.79 & 50774.73 & 0.89 & 25.5kb \\
    % & Factorjoin    & 2.73 & 11.73 & 21.91 & 54.28 & 102.98 & 1.5 & 935KB \\
    & Factorjoin    & 2.73 & 11.73 & 21.91 & 54.28 & 102.98 & 1.5 & 0.91 \\
    & TKHist(Ours)   & 1.40 & 3.84 & 6.23 & \underline{16.36} & \underline{55.04} & {10.95} & 0.77 \\
    \bottomrule
    \end{tabular}
    }
\end{table*}

Additionally, we establish time thresholds from 50ms to 600s and track query completion number within each time interval, as summarized in \Cref{fig:end:to:end:stats:time:split}.
% \dhist{}'s query completion distribution closely aligns with TrueCard's, demonstrating its superior query planning capabilities compared to other approaches.

\subsubsection*{\textbf{Q-error, latency and model size}}
We analyze Q-error, average query latency and model sizes across methods, as detailed in \Cref{tab:qerror}.
% \textcolor{green}{BEGIN FROM HERE}
% On the STATS-CEB benchmark, \dhist{} achieves the best 99th percentile and max percentile Q-error with the smallest model size of 2.7MB.
On the STATS-CEB benchmark, \dhist{} achieves the best performance in terms of the 99th percentile and maximum Q-error with the smallest model size of 2.7MB.
% For  IMDB/JOB-light, while DeepDB and BayesCard achieve the lowest Q-errors, \dhist{} delivers comparable accuracy.
For IMDB/JOB-light, while DeepDB and BayesCard achieve the lowest Q-errors, \dhist{} provides comparable accuracy.
\textbf{This demonstrates that \dhist{} produces more accurate and robust predictions than single-table statistic methods, achieving comparable or superior accuracy to learned methods.}

In terms of latency, \dhist{} achieves 10.53 ms and 10.95 ms in the STATS-CEB and IMDB/JOB-light benchmarks, respectively.
This is largely an implementation issue as \dhist{} is currently written in Python.
As demonstrated in \cref{sec:tuning}, query latency depends on hyper-parameters.
By setting the bin size to 100 and $k$ to 10, the latency reduces to 3.76 ms and 3.56 ms in the STATS-CEB and IMDB/JOB-light benchmarks, respectively, demonstrating competitive potential.

% Our end-to-end evaluation reveals that DJPD-enhanced \dhist{} achieves comparable performance to the optimal (TrueCard) while maintaining competitive query execution times.
% This highlights the robustness and efficiency of \dhist{} in real-world database systems.
% Furthermore, \dhist{}'s ability to handle frequent updates while maintaining low query latency underscores its suitability for dynamic environments.

% \begin{figure}[h!]
%     \centering
%     \includegraphics[width=1\linewidth]{fig/end_to_end_time_split.pdf}
%     \caption{End-to-end query latency distribution for STATS-CEB}
%     \label{fig:end:to:end:stats:time:split}
% \end{figure}

% To conclude, this section presents an enchanced version of \dhist{} that integrates the DJPD Algorithm.
% \dhist{} outperforms leading state-of-the-art methods by achieving significantly reduced variance in the error domain.
% \textbf{The maximum error of \dhist{} is approximately 100x, whereas other methods exhibit errors of up to 5-6 orders of magnitude.}
% In terms of other performance metrics, such as model size and query latency, \dhist{} exhibits comparable or even superior performance.

\subsection{Performance for Pure Join Queries}
\label{sec:pure:cardinality}
% \textcolor{red}{unbiased??}
In this section, we aim to produce accuracy estimation for pure join queries based on single-table statistics.
Executed without filter predicates, pure join queries specifically test the viability of using single-table statistics for join cardinality estimation. 
This isolation eliminates confounding effects from the attribute independence assumption typically required for multi-predicate selectivity estimation.
% Pure join cardinality serves as a crucial metric because it eliminates the influence of the attribute independence assumption.
Testing on pure join queries reflects the model's capability to reconstruct the actual join size using single-table statistics.
We compare \dhist{} against Join-Histogram, FactorJoin and DeepDB.
The hyper-parameters for \dhist{} in this section are set to $bin\ size$ = 200 and $k$ = 20.

\begin{table}[!ht]
    \caption{Distribution of Pure Join Queries Workload.}
        \footnotesize %scriptsize
        \label{tab:purequeries}
    \centering
    % \begin{tabular}{c|c|c|c|c|c|c}
    \scalebox{1.35}{
    % \scalebox{0.7}{
    \begin{tabular}{l|ccccc}
    \toprule
    \diagbox{Total Nums}{Join Tables} & 1 & 2 & 3 & 4 & 5 \\
    \hline
    114 & 8 & 35 & 40 & 25 & 6 \\

    \bottomrule
    \end{tabular}
    }
\end{table}

% \Cref{fig:cumulative} in the introduction section presents a boxplot illustrating the prediction accuracy for the STATS-CEB workload.
We consider all possible combinations of table joins in STATS, resulting in 114 join queries without filter predicates (see \Cref{tab:purequeries}).
% \textcolor{red}{add a worklod stats-pure??}
Those queries are categorized by the number of tables.

% \begin{figure}[t!]
%     % \vspace{-0.5cm}
%     \centering
%     \includegraphics[width=0.9\linewidth]{fig/scaling.pdf}
%     \caption{Cumulative error of join queries without filter predicates in STATS dataset. For the Workload, see \cref{sec:pure:cardinality}.}
%     \label{fig:cumulative}
%     \vspace{-0.7cm}
% \end{figure}

\begin{figure}[t!]
    \vspace{-0.4cm}
    \centering
    \includegraphics[width=0.8\linewidth]{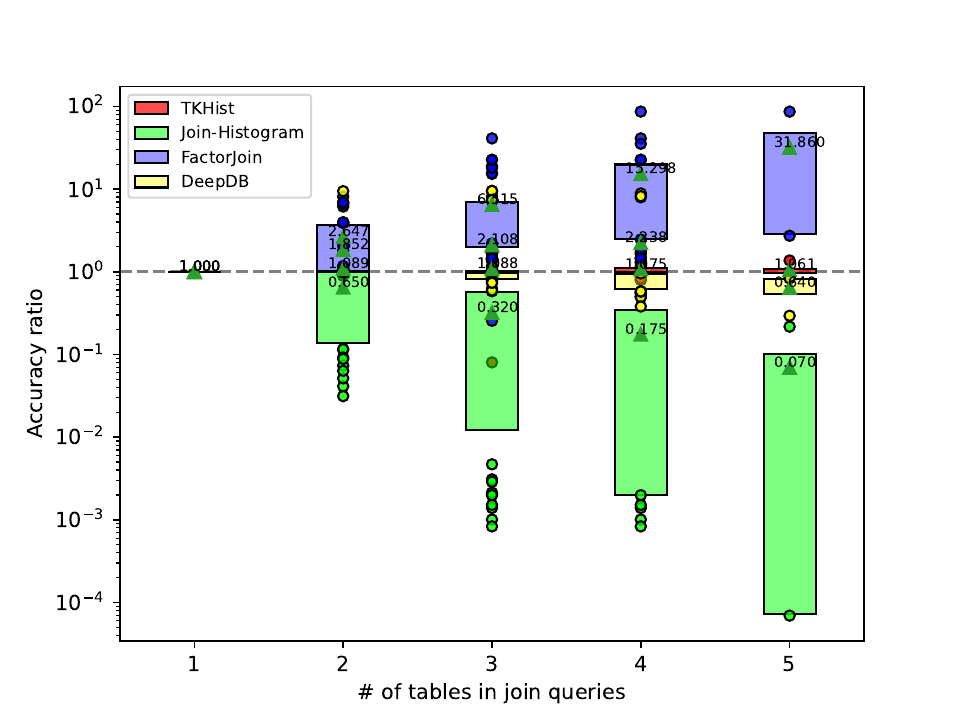}
    % \caption{Cumulative error of join queries without filter predicates in STATS dataset. For the Workload, see \cref{sec:pure:cardinality}.}
    \caption{Cumulative error of join queries without filter predicates in STATS dataset.}
    \label{fig:cumulative}
    % \vspace{-0.5cm}
\end{figure}

As \Cref{fig:cumulative} shows,
% FactorJoin tends to over-estimate and Join Histogram tends to under-estimate.
FactorJoin tends to over-estimate cardinality while Join-Histogram tends to under-estimate it,
DeepDB provides more accurate estimates but still exhibits under-estimation tendencies.
For two-table join queries, FactorJoin and Join-Histogram achieve average prediction accuracy of 65\% and 264\%, respectively.
As the number of tables increases, errors from both methods accumulate and grow exponentially.
For 5-table joins, FactorJoin's maximum error reaches 8650\% while Join-Histogram's minimum estimate drops to $6.94\times 10^{-5}$ of the actual cardinality.
% Despite lower variance, FactorJoin's errors remain substantially higher than \dhist{}'s.
In contrast, \dhist{} demonstrates superior performance with near-optimal predictions across all queries, maintaining an average error of 6.1\% for  5-table queries.
\textbf{This shows \dhist{} provides accurate cardinality estimation for pure join queries while other methods exhibit significantly higher errors.}

% \textcolor{red}{needed for bin wise ??}
\subsubsection*{\textbf{Bin-wise Accuracy Insight}}
\begin{figure}[ht!]%[htb!]
% \vspace{-0.6cm}
    \vspace{-0.5cm}
    \centering
    \includegraphics[width=0.8\linewidth]{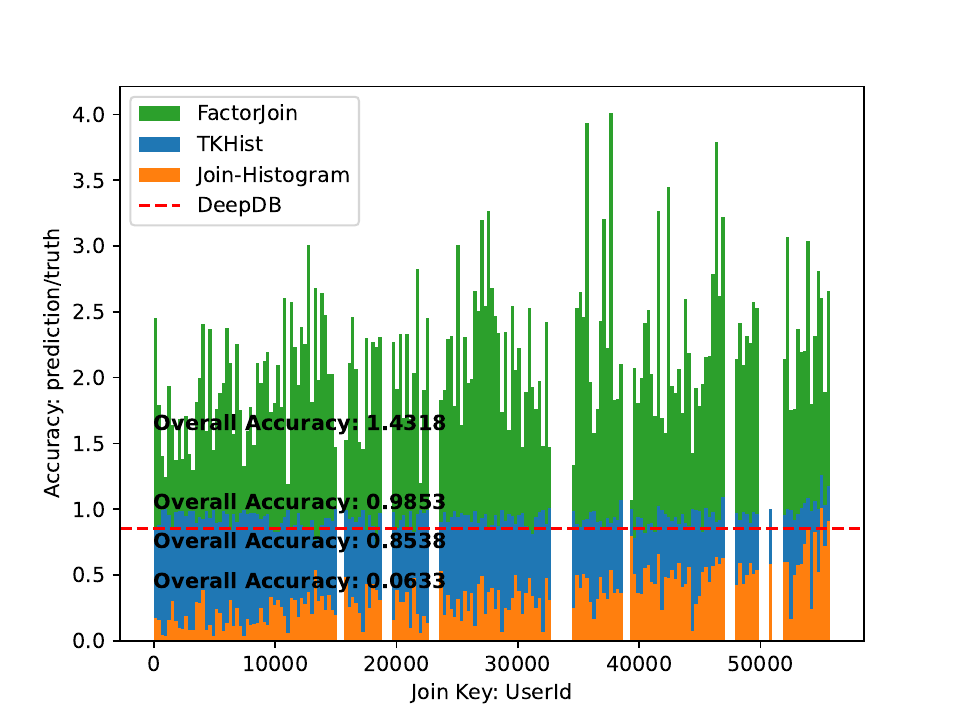}
    \caption{Bin-wise Accuracy Comparison of Cardinality Estimation Methods. The dataset is STATS, and the corresponding query is 
    $badges_{UserId}\bowtie_{UserId}comments$.
    }
    \label{fig:example-improvement}
    \vspace{-0.5cm}
\end{figure}

\Cref{fig:example-improvement} compares bin-wise estimation performance across CE methods using a representative query in STATS-CEB workload.
Join-Histogram produces severe underestimates with an overall accuracy of only 6.33\%,
FactorJoin applies the upper-bound technique to deliver a more robust prediction, achieving an accuracy of 143.18\%.
% Neither method achieves unbiased bin-level estimates.
By equipping each bin with a \topk{} container, \dhist{} achieves near-perfect predictions for each bin and produces an accuracy estimation with an overall accuracy of 98.53\%.
% \textcolor{red}{
% This also demonstrates that \dhist{} can accurately identify dominant join paths.
% }
This also demonstrates that \dhist{} can accurately identify dominant join paths.
% \textcolor{red}{unbiased}
% By equipping each bin with a \topk{} container, \dhist{} achieves near-perfect predictions for each bin with an overall accuracy of 98.53\%.
Meanwhile, DeepDB achieves an average accuracy of 85.38\%.

\subsection{Performance for Queries with Predicates}
\label{sec:multi:table:stats}
\begin{figure}[h!]
    \centering
    \includegraphics[width=1\linewidth]{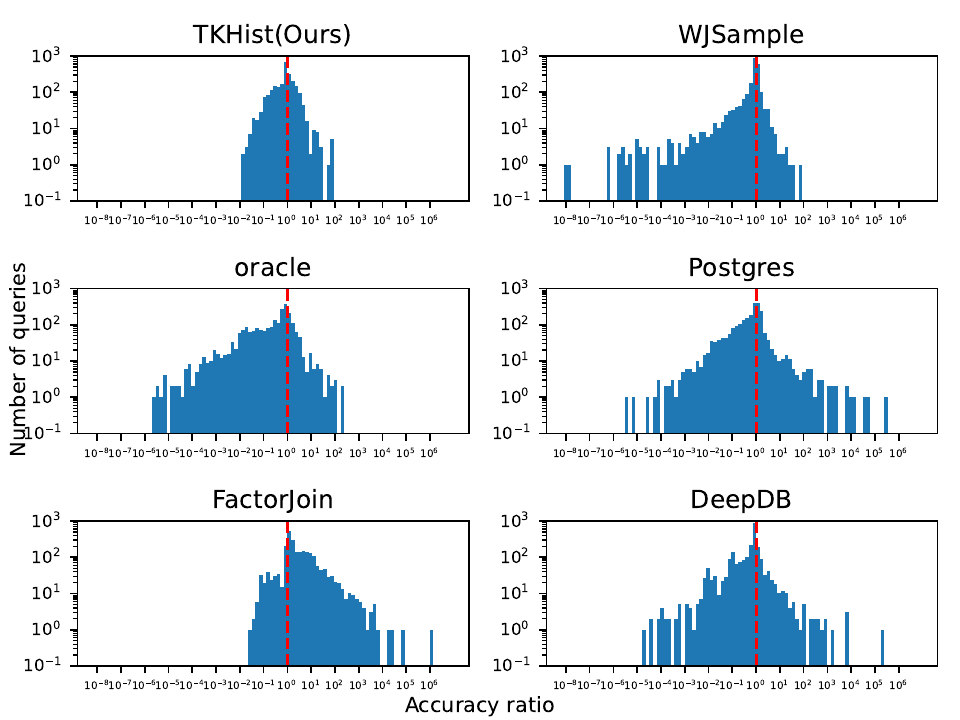}
    \caption{Accuracy ratio of multi-table queries for STATS-CEB. Ratio range from $10^{-8}$ to $10^{6}$. The red line marks 1.}
    \label{fig:multi:table:accuracy:stats}
    % \vspace{-0.5cm}
\end{figure}

In this section, we evaluates \dhist{}'s performance on real-world single-table query workloads with filter predicates.
We run experiments using the STATS-CEB benchmark \cite{han2021cardinality}, which consists of 2,603 multi-table join queries with diverse filter predicates on both categorical and continuous attributes.
As illustrated in \Cref{fig:multi:table:accuracy:stats}, \textbf{\dhist{} produces more robust join query estimation with minimal variance in the error domain, spanning 4 orders of magnitude.}
By comparison, DeepDB and FactorJoin exhibit error ranges spanning 7 orders of magnitude.
WJSample generates substantial errors for six-table joins with highly correlated filter predicates, demonstrating sampling's limitations for complex multi-join queries. 
Meanwhile, \dhist{} requires only 21 seconds for state building, significantly faster than DeepDB (1259 seconds), FactorJoin (26 seconds), and BayesCard (81 seconds).

% \subsection{\dhist{} Hyper-Parameter Tuning}
% \label{sec:tuning}
% \dhist{} has two key hyper-parameters, bin size and $k$ value, both significantly impacting system performance.
% \textcolor{red}{We use the pure join queries mentioned in \Cref{sec:pure:cardinality}.}
% % In this section, we analyze \dhist's sensitivity to these hyper-parameters.

% \subsection{Ablation Study}
\subsection{\dhist{} Hyper-Parameter Tuning}
\label{sec:tuning}
% In this section, we first show the hyper-parameter tuning of \dhist{} including $bin size$ and $k$.
% Next, we show the performance of dominant join path correlation discovery (DJPCD) algorithm.
In this section, we present the hyper-parameter tuning of \dhist{}, including $bin size$ and $k$.
By tuning these hyper-parameter, we can balance the estimation accuracy and space overhead.
% Next, we show the performance of dominant join path correlation discovery (DJPCD) algorithm.
\subsubsection*{\textbf{Bin size selection}}
\label{sec:bin:size}
We vary the bin size from 20 to 400 to evaluate its influence on \dhist{}'s performance. 
% The top-$k$ value is fixed to 10, and we repeat the same workloads using the for pure join queries.
% The top-$k$ value is fixed to 10, and we use pure join queries based on STATS dataset.
The $k$ value is fixed to 10.

\begin{figure}[h!]
    \centering
    \includegraphics[width=1.0\linewidth]{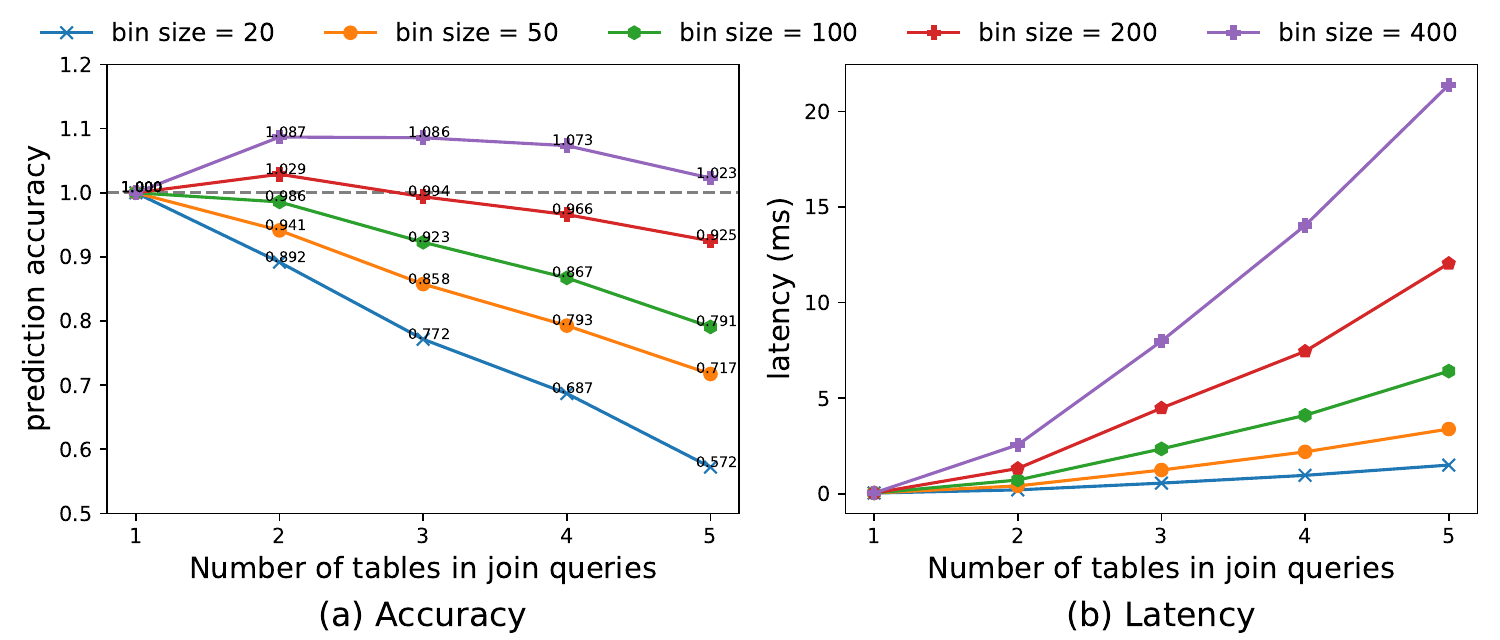}
    \caption{Tune bin size  for STATS-CEB benchmark.}
    \label{fig:tune:bin:size}
    % \vspace{-0.2cm}
\end{figure}

% As shown in \Cref{fig:tune:bin:size}.(a), even with a small bin size of 20, the overall accuracy for 5-table join query workload is 57.2\%, outperforming  Join-Histogram and FactorJoin (see \Cref{fig:cumulative}).
% For bin sizes $\geq$ 50, the overall prediction accuracy ranges from 71.7\% to 108.7\%.
% As the number of tables in join queries increases, the overall accuracy decreases slightly.
% \dhist{} achieves the best accuracy when bin size is 200.
% At a bin size of 400, a slight over-estimation of pure cardinality is observed.
% This error results from the approximation of using background average value (BAC) to compensate for top-$k$ join key multiplication (Line 8-13 in \Cref{alg:join:dhist}).
As shown in \Cref{fig:tune:bin:size}.(a), with a small bin size of 20, the accuracy for the 5-table join query workload is 57.2\%, outperforming Join-Histogram and FactorJoin (see \Cref{fig:cumulative}). For bin sizes $\geq$ 50, the accuracy ranges from 71.7\% to 108.7\%. 
As the number of tables increases, accuracy slightly decreases. 
\dhist{} achieves the best accuracy at a bin size of 200.
At a bin size of 400, slight over-estimation occurs due to the use of background average value (BAC) to adjust for top-$k$ join key multiplication (lines 8-13 in \Cref{alg:join:dhist}).

\Cref{fig:tune:bin:size}.(b) illustrates  bin size's impact. 
It is evident that the query latency increases as the bin size grows.
For a bin size of 100 or 200, the query latency for a 5-table join query is approximately 5 ms and 12 ms, respectively.
As query latency is a critical factor, a bin size exceeding 200 is not recommended.
The space overheads range from 0.5 to 3 MB, and we omit it for space reasons.

\subsubsection*{\textbf{k value selection}}
\label{sec:k}
This section examines the impact of the $k$ value on accuracy, latency, and space overhead, with the bin size fixed at 100 for fair comparison.

\begin{figure}[ht!]
    \centering
    \includegraphics[width=1.0\linewidth]{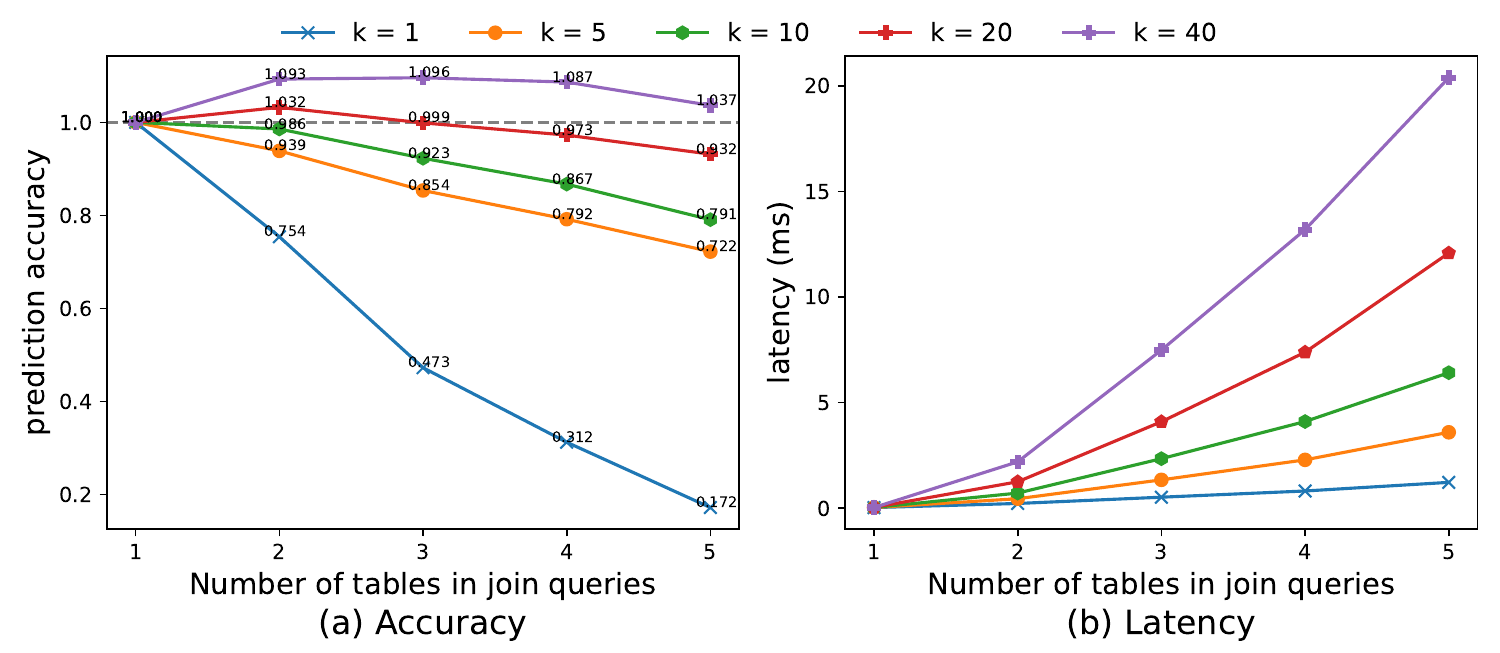}
    \caption{Tune $k$ for STATS-CEB benchmark.}
    \label{fig:tune:k}
    \vspace{-0.2cm}
\end{figure}

% \Cref{fig:tune:k}.(a) demonstrates the accuracy of \dhist{} across different $k$ values.
% When $k=0$, \dhist{} degrades to Join-Histogram, with pure cardinality estimation accuracy declining as tables increase.
% Even with $k=1$, \dhist{} achieves significantly  higher accuracy for 5-table joins compared to Join-Histogram and FactorJoin (see \Cref{fig:cumulative}).
% Typically, a $k$ value between 10 and 20  generally offers optimal results.

% \Cref{fig:tune:k}.(b) demonstrates the effect of $k$ on query latency. 
% Obviously, latency grows as $k$ increases.  
% When k = 10 (20), the query latency is ca. 7ms (11ms) for a 5-table join query. 
% The space overhead is ca. 1.4MB (2MB) when $k$ is set as 10 (20), which is quite small compared with the actual data.
\Cref{fig:tune:k}.(a) demonstrates the accuracy of \dhist{} across different $k$ values.
When $k=0$, \dhist{} degrades to Join-Histogram, with accuracy decreasing as tables increase.
At $k=1$, \dhist{} significantly outperforms Join-Histogram and FactorJoin for 5-table joins (see \Cref{fig:cumulative}).
% Typically, a $k$ value between 10 and 20 generally offers optimal results.

\Cref{fig:tune:k}.(b) demonstrates the effect of $k$ on query latency. 
Obviously, latency increases as $k$ increases.  
When k = 10 (20), the query latency is ca. 7ms (11ms) for a 5-table join query. 
The space overhead is ca. 1.4MB (2MB) when $k$ is set to 10 (20), which is quite small compared to the actual data.

The sensitivity analysis reveals that \dhist{} can be optimized through  careful tuning  of  bin size and top-$k$ value. 
% A bin size of 100-200 combined with $k$ value of 10-20 achieves the best accuracy-resource balance for efficient pure cardinality estimation.
A bin size of 100-200, combined with a $k$ value between 10 and 20, achieves the best balance between accuracy and resource efficiency for pure cardinality estimation.
% Unless otherwise specified, we use $k$ = 20 and $bin\ size$ = 200 in subsequent experiments.

% \subsubsection*{\textbf{Performance of Dominant Join Path Correlation}}
% \subsection{\textbf{Correlation Effect of Top-k Join Paths }}
\subsection{\textbf{Correlation Effect of Dominant Join Paths }}
% \subsection{Ablation}
In this section, we present the performance of the Dominant Join Path Correlation Discovery (DJPCD) algorithm.
As discussed in the motivation section, \topk{} join paths typically dominate the final join results, and correlations between filter predicates and join paths can introduce significant estimation errors. 
To address this challenge,  we employ the Dominant Join Path Correlation Discovery (DJPCD) algorithm described in \Cref{sec:dominating:join:key:discovery} to manage correlations between join keys and  filter predicates.
The workload here is the same as the multi-table queries workload in \cref{sec:multi:table:stats}.
% \textcolor{red}{...}

% We run experiments for the STATS-CEB benchmark, which comprises 2,603 multi-table join queries with diverse filter predicates on both categorical and continuous attributes.

\begin{figure}[ht!]
    % \vspace{-0.5cm}
    \vspace{-0.4cm}
    \centering
    \includegraphics[width=0.8\linewidth]{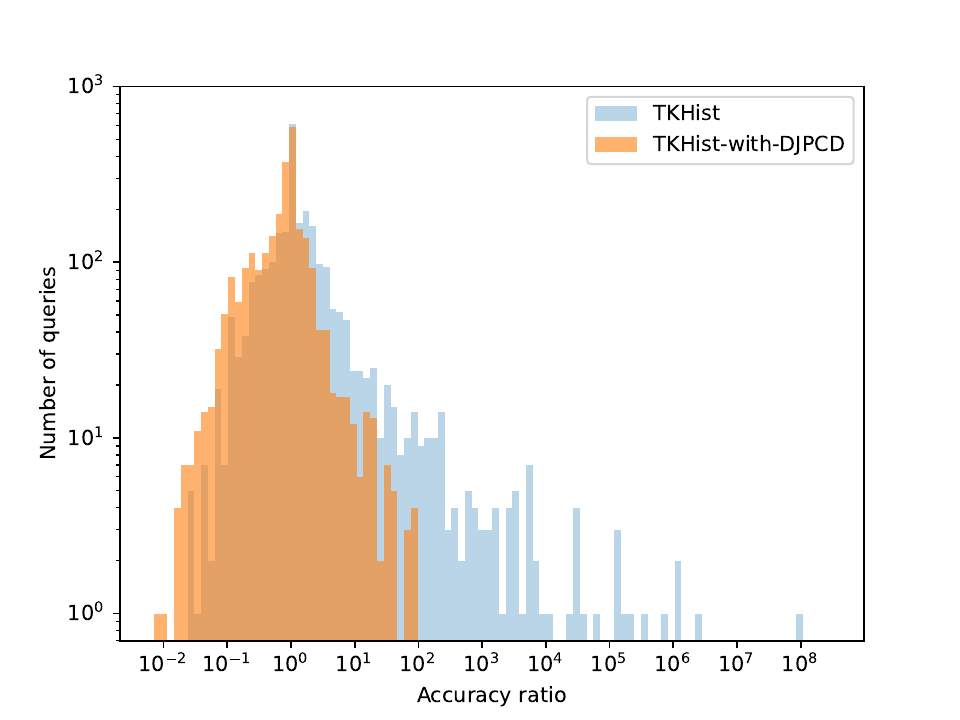}
    \caption{Correlation  effect between join paths \& predicates}
    \label{fig:correlation:effect}
    \vspace{-0.5cm}
\end{figure}

\Cref{fig:correlation:effect} compares the prediction accuracy of \dhist{} with and without the DJPCD algorithm.
The relative error range of \dhist{} without DJPCD spans from $10^{-2}$ to $10^8$, rendering it impractical for real-world deployment. 
While \dhist{} provides unbiased cardinality estimation under ideal conditions, its dependence on the independence assumption between selection attributes across tables introduces substantial errors. 
This issue is exacerbated when filter predicates eliminate dominant \topk{} join paths, leading to order-of-magnitude over-estimations.
The DJPCD algorithm effectively mitigates these issues by explicitly accounting for critical join path correlations. 
As shown in \Cref{fig:correlation:effect}, implementing DJPCD reduces the error range between $10^{-2}$ to $10^2$, demonstrating our method's practical utility for complex multi-table queries.
% \vspace{-0.3cm}

% \vspace{-0.5cm}
\section{Related Work}
We briefly review the literature on learned \card{} methods.
Initial query-driven \card{} methods use lightweight machine learning models \cite{dutt2019selectivity,park2020quicksel} to map the relationship between query workloads and actual cardinality.
Recent efforts apply neural networks for this purpose \cite{kipf2019estimating,kipf2018learned,liu2021fauce,sun2019end}.
ALECE \cite{li2023alece} discovers the implicit relationships between queries and data using attention mechanisms. 
\cite{zeighami2023neurosketch} uses NeuroSketch to provide approximate answers to single-table range aggregate queries.
In case of data and workload drifts, Warper \cite{li2022warper} adapts different \card{} models for single-table and join queries, and is able to update \card{} models using different queries.
Query-driven methods usually achieve good estimation with low latency.
However, they mainly excel for popular/predictable workloads as training data must be prepared in advance. 

State-of-the-art multi-table data-driven \card{} methods mainly fall into two categories: 1) methods that use de-normalized join results or fanout techniques to build sophisticated large/deep models to fit the high dimensional data distribution \cite{getoor2001selectivity,hilprecht2020deepdb,tzoumas2011lightweight,wu2020bayescard,yang2020neurocard,zhu2021flat,kim2022learned}.
However, they are difficult to deploy in existing database systems for a fair end-to-end comparison, and their accuracy deteriorates with an increasing number of tables in join queries.
2) Methods that rely merely on single-table statistics. They are easy to deploy with low maintenance cost, more friendly to updates, and capable of producing more robust query plans. 
FactorJoin \cite{wu2023factorjoin} and SafeBound \cite{deeds2023safebound} use information theory to produce an upper bound for \card{}.
ASM \cite{kim2024asm} combines single-table autoregressive models and sampling to produce more accurate estimation for join queries.

In addition, studies based on single-table statistics are closest to our work, including Join-Histograms \cite{adellera}, FactorJoin \cite{wu2023factorjoin}, ASM \cite{kim2024asm}, top frequency histograms \cite{oracleHistogram}, etc.
% The top frequency histogram is adopted in Oracle database systems to ignore unpopular values that are statistically insignificant. However, the cumulative influence of dominating items, and the correlation between join keys and filter predicates are not studied in depth.

\vspace{-0.5cm}
\section{Conclusion}
In this study, we introduce a novel histogram structure, \dhist{}, that incorporates a top-$k$ join key container to capture non-uniformity within bins.
Within the current binary-join framework, errors from state-of-the-art methods accumulate and grow exponentially as the number of tables in the query increases.
\dhist{} provides accurate cardinality estimation for multi-table pure join queries without filter predicates.
Moreover, we found that the attribute independence assumption leads not only to underestimation but also to overestimation in multi-table join queries. 
The correlation between join keys and filter predicates is stronger than the correlation among filter predicates themselves.
\dhist{} addresses these correlations through the dominating join key correlation discovery algorithm and outperforms state-of-the-art methods in query accuracy, achieving a 2-3 order of magnitude reduction in the error domain.
Future work includes utilizing better data structures, such as Bloom Filters, to optimize dominating join key discovery, applying \dhist{} to database systems, developing improved binning strategies, and exploring more efficient join strategies between \dhist{}s.

\vspace{-5pt}
\section*{Acknowledgments}
This work is supported by projects funded by Priority Academic Program Development of Jiangsu Higher Education Institutions and the Huawei "Smart Base" Industry-Education Integration Collaborative Education Program.

\section*{GenAI Usage Disclosure}
In all stages of our work, generative AI tools are not employed to produce code or data.
The author confirms that all scientific arguments and data analyses reflect their own intellectual contributions.
Generative AI was used solely to improve the quality of text, including spelling, grammar, punctuation, clarity, etc.
This limited use was aimed at enhancing the clarity and readability of the text, in accordance with academic integrity standards.

\balance

\end{document}